\documentclass[twocolumn]{aastex701}
\usepackage{amsmath}

\usepackage[T1]{fontenc} 

\let\tablenum\relax 
\usepackage[print-unity-mantissa=false, uncertainty-mode=separate, range-phrase=--]{siunitx}
\DeclareSIUnit\parsec{pc} 
\DeclareSIUnit\year{yr}
\DeclareSIUnit\arcsecond{as} 
\DeclareSIUnit\dex{dex}

\DeclareRobustCommand{\okina}{%
  \raisebox{\dimexpr\fontcharht\font`A-\height}{%
    \scalebox{0.8}{`}%
  }%
}

\newcommand{\OO}{\={O}tautahi--Oxford model}

\newcommand{\FeH}{[\mathrm{Fe}/\mathrm{H}]}
\newcommand{\aFe}{[\alpha/\mathrm{Fe}]}

\submitjournal{AJ}

\shorttitle{ISO Compositions}
\shortauthors{Hopkins et al.}

\begin{document}

\title{On the Compositions of Interstellar Objects}

\correspondingauthor{Matthew J. Hopkins}
\email{matthew.hopkins@canterbury.ac.nz}
\author[0000-0001-6314-873X]{Matthew J. Hopkins}
\affiliation{School of Physical and Chemical Sciences --- Te Kura Mat\={u}, University of Canterbury, Private Bag 4800, Christchurch 8140, New Zealand}
\email{matthew.hopkins@canterbury.ac.nz}

\author[0000-0003-3257-4490, gname=Michele, sname=Bannister]{Michele T. Bannister}
\affiliation{School of Physical and Chemical Sciences --- Te Kura Mat\={u}, University of Canterbury, Private Bag 4800, Christchurch 8140, New Zealand}
\email{michele.bannister@canterbury.ac.nz}

\author[0000-0002-8868-7649]{Bertram Bitsch}
\affiliation{Department of Physics, University College Cork, Cork, Ireland}
\email{bbitsch@ucc.ie}

\author[0000-0001-5578-359X, gname=Chris, sname=Lintott]{Chris Lintott}
\affil{Department of Physics, University of Oxford, Denys Wilkinson Building, Keble Road, Oxford, OX1 3RH, UK}
\email{chris.lintott@physics.ox.ac.uk}

\author[0000-0002-1975-4449]{John C. Forbes}
\affiliation{School of Physical and Chemical Sciences --- Te Kura Mat\={u}, University of Canterbury, Private Bag 4800, Christchurch 8140, New Zealand}
\email{john.forbes@canterbury.ac.nz}

\begin{abstract}  
As physical samples of distant planetary systems, the compositions of interstellar objects (ISOs) should correlate with the properties of their diverse parent stars.
We combine a chemical partition model with a sample of stellar elemental abundances from APOGEE DR17 to predict the abundances of commonly-observed cometary volatiles in ISOs.
We find that the compositions of ISOs vary significantly between different stars.
In particular, we predict a strong correlation between an ISO's ammonia abundance and its parent star's metallicity: high-metallicity stars will create ISOs rich in ammonia.
This suggests the production rates of NH$_3$'s photolytic daughter products are an ideal observational tracer for the origins of ISOs.
We infer that 2I/Borisov formed around a star of near-solar metallicity ($-0.4\lesssim\FeH\lesssim0.3$), and corroborate the velocity- and isotope-based inferences that 3I/ATLAS formed around a lower-metallicity star ($-0.8\lesssim\FeH\lesssim0.0$).
The production rates of 2I and 3I imply both only contain hypervolatiles such as CO and N$_2$ that are trapped in other less-volatile ices, similar to typical Solar System comets.
Despite this, we argue that true hypervolatile-ice-rich ISOs may exist; they would exhibit high CO and N$_2$ production rate ratios, possibly similar to those of C/2016 R2 (PanSTARRS).
Our predictions provide context for future ISO discoveries, opening a path to link the properties of their origin systems to observable production rates.

\end{abstract}


\section{Introduction}

The interstellar objects (ISOs) we see passing through the Solar System formed in the protoplanetary disks of distant stars, and possibly in very different conditions to the Solar System's comets.
Because a star and its planetesimals, its eventual ISOs, form from the same collapsing molecular cloud core \citep{Oberg2021}, the composition of an ISO is linked to the elemental abundances of its parent star.
These elemental abundances are regularly observed in large Milky Way stellar surveys; they show wide variance and correlations with age and metallicity due to the complex chemical evolution of the Galaxy \citep[e.g.][]{Weinberg2019}.
This presents the possibility of predicting the distribution of compositions in the Galactic ISO population from the stellar population. 

Cometary activity is driven by both the sublimation of volatile ices (e.g. H$_2$O and CO$_2$) as well as the release of hypervolatiles trapped in these ices (e.g. CO) \citep{Meech2009,Womack2017}.
The production rates of these species from a comet is measured by spectroscopy of its coma, and their ratios permit the composition of a cometary body to be inferred \citep{Biver2024}.
2I/Borisov and 3I/ATLAS are the only two cometary ISOs observed so far, but they already show a remarkable diversity in properties compared to each other and to Solar System comets, indicative of their formations around different parent stars \citep{Fitzsimmons2024}.
Velocity- and isotope-based methods indicate 3I is likely to be far older than both 2I and the Solar System, and to have formed around a lower-metallicity star \citep{Hopkins2025a,Taylor2025,Cordiner2026,Opitom2026}.
1I/\okina Oumuamua had no visible cometary activity, though its non-gravitational acceleration implies low-level outgassing of some volatile \citep{Micheli2018}.

While the chemistry in protoplanetary disks and planetesimal evolution is undoubtedly complex, first-order models can be informative as a starting point.
The \OO{}\footnote{Available at \url{https://github.com/Otautahi-Oxford/OO-model}}  \citep{Lintott2022,Hopkins2023,Hopkins2025b,Hopkins2025a,Dorsey2025,Forbes2025} predicts properties of the Milky Way's ISO population from the stellar population; previous iterations include water mass fraction, assumed to depend only on parent star metallicity, as a basic proxy for ISO composition.

Here we investigate ISO composition in significantly more detail, predicting trends for a wider range of volatiles formed from C, N, O and S in cometary ISOs, informed by the species commonly observed in Solar System comet comae \citep[e.g. Table 3 of][]{Biver2024}.
We construct a chemical model linking the molecular compositions of ISOs to the elemental compositions of their parent stars (\S\ref{sec:ppd} \& \S\ref{sec:ppd2iso}), then apply this to a well-characterised sample of Milky Way stars (\S\ref{sec:apogee}) to predict and investigate the compositions of the Galactic population of ISOs (\S\ref{sec:results}).
In \S\ref{sec:2I3I} we compare the observed production rate ratios of 2I/Borisov and 3I/ATLAS to Solar System comets (summarised in Appendix~\ref{sec:prodrates}), demonstrating that 2I originated around a solar-metallicity star and 3I originated around a low-metallicity star.
In \S\ref{sec:hypervolatiles} we discuss the existence of hypervolatile-ice-rich ISOs which contain CO and N$_2$ as pure ices, as opposed to hypervolatiles trapped in other less-volatile ices.
Finally, we explore the implications of this work for planetary systems and the Galaxy (\S\ref{sec:implications}), and in \S\ref{sec:future} we summarise the observations required to make the most of future ISO discoveries.

\section{Methods}

\subsection{Stellar Sample: APOGEE DR17}\label{sec:apogee}
APOGEE \citep{Majewski2017} is a ground-based spectroscopic stellar survey; its infrared wavelength range and high signal-to-noise ratio allows the accurate measurement of the key elemental abundances we require (C, N, O) via molecular lines \citep{GarciaPerez2016}.
We access the APOGEE data in SDSS DR17 \citep{Abdurro'uf2022} via the \texttt{allStar} file, which contains calibrated relative element abundances \texttt{X\_FE}. 

We include in our sample only those stars which have measured abundances for all the elements we require: namely [C/Fe], [N/Fe], [O/Fe], [Mg/Fe], [Si/Fe], [S/Fe] and [Fe/H] for the chemical model, and additionally [Ca/Fe] for the calculation of each star's [$\alpha$/Fe] abundance. 
We remove giants with a \(\log g>3.8\) cut, as these stars do not have surface elemental abundances representative of their natal compositions, due to dredge up \citep{Smith1985} and planetary engulfment \citep{Carlberg2012}.
Finally, following \cite{Holtzman2015}, we cut to only APOGEE stars with the bitmask \texttt{EXTRATARG}=0 to remove special targets outside of the main red star survey, trimming to a homogeneous sample.
This leaves us with a sample of 85,775 stars from a broad swathe of the Milky Way, representative of the range of compositions in the Galaxy's stellar population.

\begin{figure*}
    \centering
    \includegraphics[width=0.5\linewidth]{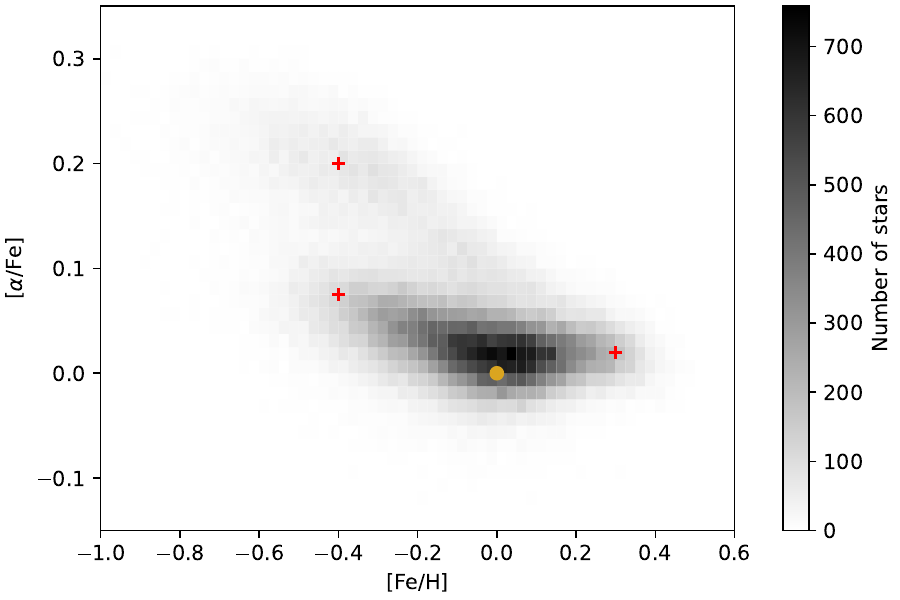}
    
    \includegraphics[width=0.48\linewidth]{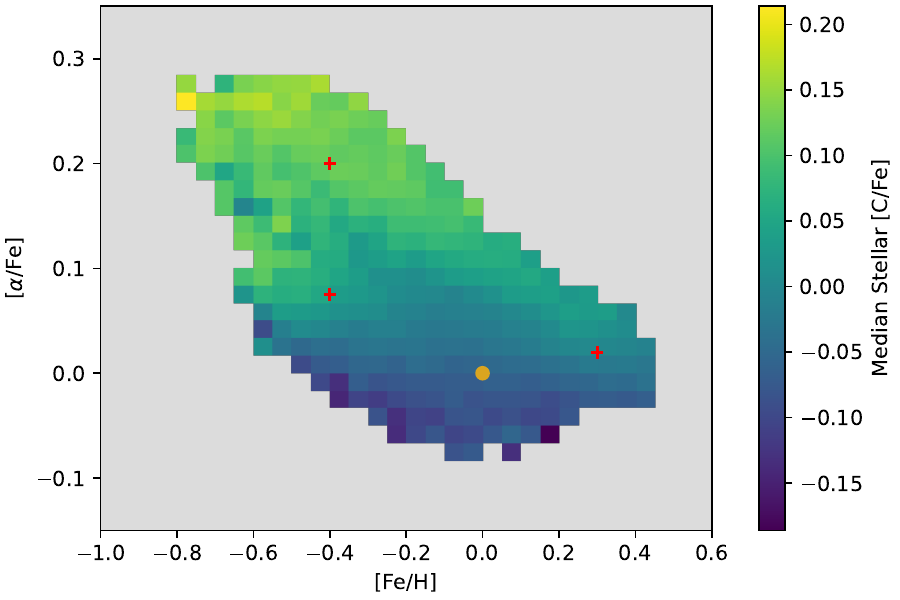}
    \includegraphics[width=0.48\linewidth]{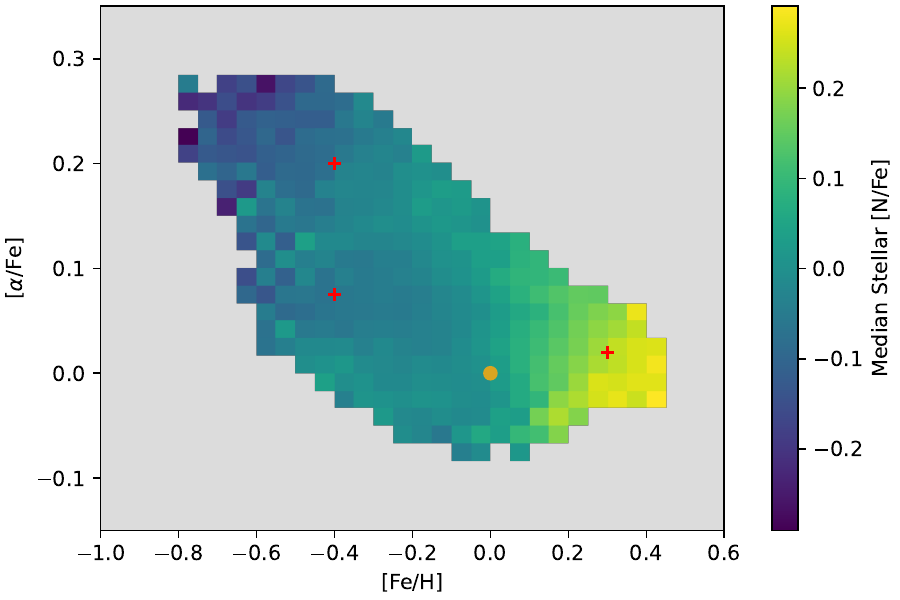}
    
    \includegraphics[width=0.48\linewidth]{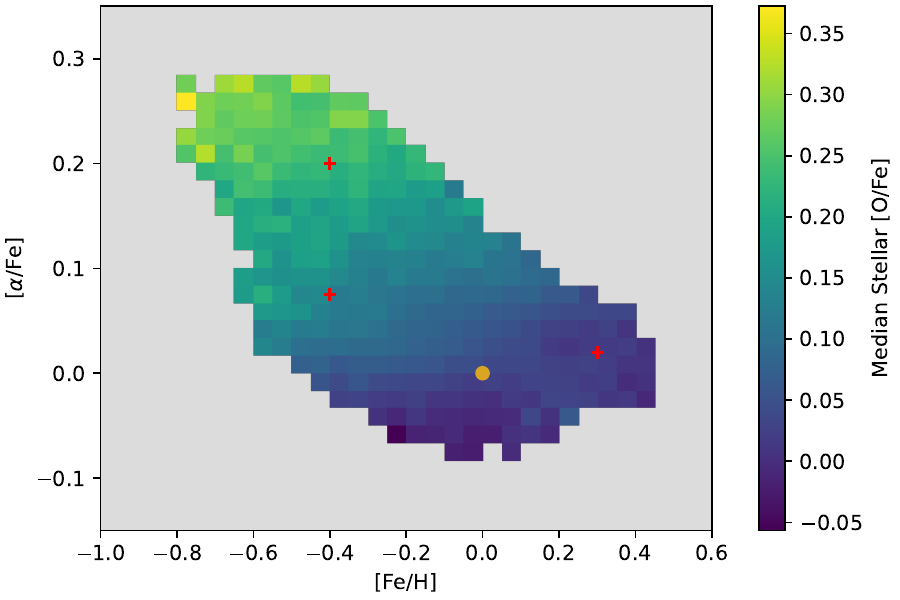}
    \includegraphics[width=0.48\linewidth]{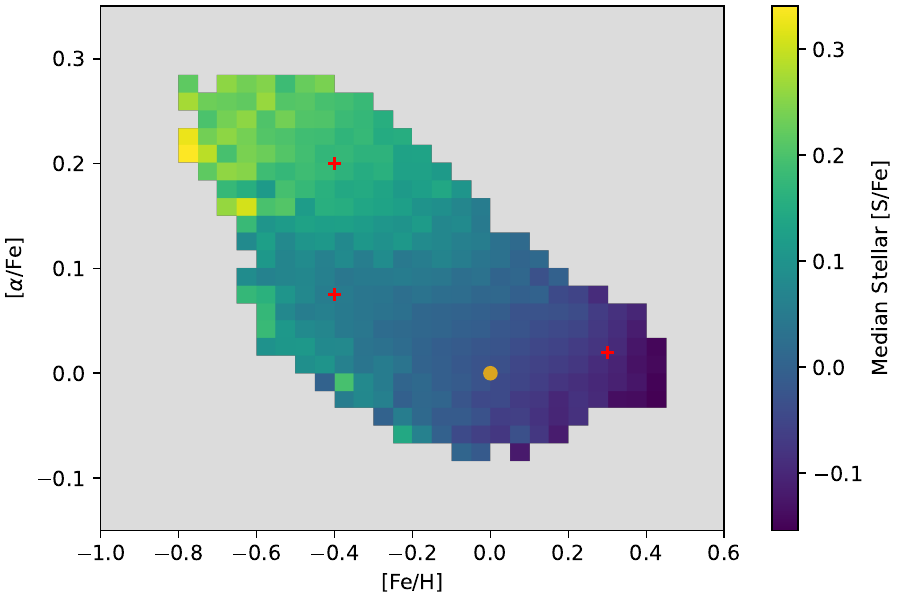}

    \caption{Plot of the distribution of our APOGEE sample in [Fe/H]--[$\alpha$/Fe] space, and the median stellar abundances of the major elements for cometary volatiles. The gold circle marks the position of the Sun in this space, and red crosses mark points at high [Fe/H], low [Fe/H], and high $\aFe$ at which the distribution of abundances is plotted in Fig~\ref{fig:aFeH_points}.}
    \label{fig:stars}
\end{figure*}
To investigate chemical trends we distribute our stellar sample by [Fe/H] and $[\alpha/\mathrm{Fe}]=([\mathrm{Mg}/\mathrm{Fe}] + [\mathrm{Si}/\mathrm{Fe}] + [\mathrm{Ca}/\mathrm{Fe}])/3$.\footnote{[Ti/Fe] is another alpha element used in other works, but it is deemed `deviant' in dwarfs in APOGEE DR17. See \url{https://www.sdss4.org/dr17/irspec/abundances/}}
Plotted in Fig.~\ref{fig:stars} (top) is the Tinsley--Wallerstein diagram \citep{Buder2021} of our sample; as expected, the majority of stars lie on a non-$\alpha$-enhanced sequence with a wide range of metallicities, with a minority population on an $\alpha$-enhanced sequence extending to low metallicity \citep{Tinsley1979}.
Also plotted in Fig.~\ref{fig:stars} are the median stellar abundances of [C/Fe], [N/Fe], [O/Fe], and [S/Fe] in bins in [Fe/H]--[$\alpha$/Fe] space containing at least 20 stars.
These elements form the most common volatile molecules observed in comets \citep[cf. Fig.~9 of][]{Biver2024}. 
Importantly, nitrogen shows the opposite trend to the other elements, increasing strongly with [Fe/H].

\subsection{Protoplanetary Disk Chemical Model}\label{sec:ppd}

\begin{table*}
\centering
\begin{tabular}{c|c|c|c}
Species ($X$) &$T_\mathrm{cond.}\,/\,\mathrm{K}$ & Volatility &Relative Abundance of Species $X$ ($N_\mathrm{X}$) \\\tableline
CO &$20$ & Hypervolatile & 0.45 $\times N_\mathrm{C}$  \\
N$_2$ & $20$ &Hypervolatile& \(0.45\times N_\mathrm{N}\)\\
CO$_2$  &70 &Volatile&0.05 $\times N_\mathrm{C}$\\
NH$_3$ &90 &Volatile& \(0.02\times N_\mathrm{N}\)\\
H$_2$S &150 &Volatile& \(0.1\times N_\mathrm{S}\) \\
Refractory carbon &&Refractory& 0.50 $\times N_\mathrm{C}$\\
Refractory nitrogen &&Refractory& \(0.08\times N_\mathrm{N}\)\\
FeS &&Refractory&\(0.9\times N_\mathrm{S}\) \\
Fe$_3$O$_4$ &&Refractory& \(\frac{1}{6}\times( N_\mathrm{Fe} -  N_\mathrm{FeS})\) \\
Fe$_2$O$_3$ &&Refractory&  \(\frac{1}{4}\times( N_\mathrm{Fe} -  N_\mathrm{FeS})\) \\
Mg$_2$SiO$_4$ &&Refractory& \(\max(N_\mathrm{Mg}-N_\mathrm{Si},\,0)\)\\
SiO$_2$ &&Refractory& \(\max(N_\mathrm{Si}-N_\mathrm{Mg},\,0)\)\\
MgSiO$_3$ &&Refractory& \( N_\mathrm{Mg}-2\times N_{\mathrm{Mg}_2\mathrm{SiO}_4}\)\\[0.2cm]
H$_2$O &150 &Volatile& 
$\begin{aligned}
N_\mathrm{O}-(& N_\mathrm{CO} + 2\times N_{\mathrm{CO}_2} + 4\times N_{\mathrm{Fe}_3\mathrm{O}_4} \\
&+ 3\times N_{\mathrm{Fe}_2\mathrm{O}_3}  +4\times N_{\mathrm{Mg}_2\mathrm{Si}\mathrm{O}_4} \\
&+3\times N_{\mathrm{Mg}\mathrm{Si}\mathrm{O}_3} + 2\times N_{\mathrm{Si}\mathrm{O}_2})
\end{aligned}$\\
\end{tabular}
\caption{Partition model giving the relative abundances of molecules in the protoplanetary disk of a star with relative elemental abundances $N_\mathrm{X}=10^{[\mathrm{X}/\mathrm{Fe}]+[\mathrm{Fe}/\mathrm{H}]+[\mathrm{X}/\mathrm{H}]_\odot}$. Condensation temperatures $T_\mathrm{cond.}$ from \cite{Schneider2021}.}
\label{tab:chem}
\end{table*}

As a molecular cloud core collapses, the star at its centre forms from the same material as the protoplanetary disk around it \citep[e.g.][]{Oberg2021}. 
This protoplanetary disk evolves into a planetary system which ejects interstellar objects shortly after formation \citep{Brasser2006}, but the star will largely retain the same surface elemental abundances throughout its life on the main sequence.
Therefore the molecular composition of a star's protoplanetary disk, and thus the composition of the ISOs it produced, can be predicted from the star's present-day observed elemental abundances \citep{Huhn2023}.

We do this with the partition model specified in Table~\ref{tab:chem}.
This sequentially distributes the available atoms of each element into different species with the partition fractions given in the last column, estimating the abundances of molecules in the protoplanetary disk of each star in our APOGEE sample.
We mark species as ``volatile'' and ``refractory'' as a loose distinction between species that will or will not be sublimating from cometary ISOs at the typical heliocentric distances at which they are observed, $1\,\mathrm{au}\lesssim r_h \lesssim5\,\mathrm{au}$.
``Hypervolatiles'' are species with very low sublimation temperatures ($\sim20\,\mathrm{K}$), and therefore are only present in typical Solar System comets when trapped in other ices \citep[][]{Meech2009}.

This model takes the list of species formed and the partition order from the models of \cite{Bitsch2020}, \cite{Schneider2021}, and \cite{Cabral2023}, with the partition fractions adapted for predicting cometary ISO compositions.
We avoid overfitting to Solar System comets by deriving the partition fractions from standard results in studies of the interstellar medium and protoplanetary disks in general \citep[e.g.][]{Bergin2015,Krijt2023} rather than choosing them so that the model is required to reproduce the compositions of Solar System comets; we can then apply the model to the Sun's elemental abundances \citep{Grevesse2007} and use the broad agreement of our model's predictions with observations of Solar System comets to check its validity.

Approximately half of the carbon in the ISM is in a refractory form \citep{Cardelli1996}, with the majority of the rest in CO \citep[][]{Lewis1980} and the remainder forming CO$_2$, so we split the carbon between these in a 50:45:5 ratio.
The dominant form nitrogen takes is N$_2$ \citep[][assumed 90\%]{Lewis1980,Owen2001}, with the remainder split between volatile NH$_3$ and a refractory nitrogen reservoir in a 1:4 ratio based on the findings of \cite{Altwegg2020}.
This refractory nitrogen represents the separate source of NH$_3$ released when comets reach extremely low heliocentric distances \citep[$r_h<1\,\mathrm{au}$;][]{DelloRusso2016}, which may be a form of ammonium salts \citep{Poch2020}. 
We treat this refractory nitrogen species as a separate species to volatile NH$_3$ because the majority of discoverable ISOs will not reach the low heliocentric distances required for its sublimation \citep{Dorsey2025}.
We assume 90\% of sulfur forms refractory FeS \citep{Kama2019} and assume that the rest takes the form of H$_2$S as this is the majority volatile carrier of sulfur \citep{Calmonte2016}. 
The model's use of fixed partition fractions vastly simplifies the complex astrochemistry which occurs in both the pre-stellar core and protoplanetary disk.
However, the most important features of this model are the trends it predicts between stellar and ISO abundances, and these are robust.

A small number of stars in our sample fall outside the range of validity of this recipe: 2487 with $N_\mathrm{Fe}<0.9\times N_\mathrm{S}$ are assigned a negative abundance of iron oxides, and 1028 have so little oxygen relative to other elements that a negative $N_{\mathrm{H}_2\mathrm{O}}$ is assigned.
In the low-oxygen regime, the deficit would plausibly be made up by refractory species taking less oxygen-rich forms, e.g. pure iron instead of iron oxides.
We do not explicitly model the chemistry in these regimes, and since the number of stars affected is small we remove them.
There remains an interesting possibility for future modelling of the rare and chemically-unusual ISOs that may originate in protoplanetary disks at the edges of this chemical space.

\subsection{Deriving ISO Composition}\label{sec:ppd2iso}

We assume that cometary ISOs largely inherit their compositions directly from the protoplanetary disk compositions predicted by the model above.
This is a good assumption for all of the non-hypervolatile species listed in Table~\ref{tab:chem}: in the cold outer regions of a protoplanetary disk these species are entirely in solid states, and therefore are incorporated into forming cometary bodies with the ratio as in the wider protoplanetary disk.

However, this does not apply to hypervolatiles.
Typical $\sim1\,\mathrm{km}$-diameter solar system comets do not contain CO or N$_2$ ice; their condensation temperatures ($\sim20\,\mathrm{K}$) are significantly lower than the estimated formation temperatures of Solar System comets \citep[$30\textrm{--}50\,\mathrm{K}$;][]{Rubin2020}, and any CO or N$_2$ ice originally incorporated into a typical km-scale comet would have been driven off by a range of heating mechanisms that act on both long- and short-period comets \citep{Stern2003,Mousis2012,Davidsson2021,Guilbert-Lepoutre2024,Parhi2026}.
The CO and N$_2$ observed in the comae of these comets is not produced by the sublimation of CO and N$_2$ ices, but instead is due to the release of CO and N$_2$ trapped in other less-volatile ices as they undergo phase transitions \citep{Bar-Nun1985,Meech2009,Womack2017,Simon2019}.
There are exceptions, for example C/2016 R2 which is thought to have a composition with significant CO and N$_2$ ices \citep{Mousis2021,Lisse2022}, however these are rare amongst Solar System comets. 
2I and 3I both appear to also only contain hypervolatiles trapped in other ices, based on the comparative similarity of their $Q(\mathrm{CO})/Q(\mathrm{H}_2\mathrm{O})$ production rate ratios to solar system comets at the same heliocentric distances (Appendix~\ref{sec:prodrates}), as well as 3I's coma N$_2^+$/CO$^+$ abundance \citep{Ferellec2026}.

The abundances of CO and N$_2$ in the bulk nucleus compositions of these comets is not set by the composition of the protoplanetary disk; instead, their abundances are set by the trapping efficiencies of each hypervolatile with each host ice. 
Because the host ice is not certain \citep[amorphous water ice, crystalline water ice and CO$_2$ ice are all possibilities just within the Solar System;][]{Simon2019,Prialnik2024} we do not predict the abundances of hypervolatiles in ISOs here.
We discuss the existence of hypervolatile-ice-rich ISOs and make appropriate predictions in \S\ref{sec:hypervolatiles}.

The molecules we have included in our principal predictions constitute the major volatile species found in comets, and the abundance of each species has a clear dependence on a corresponding parent star elemental abundances, e.g. ISO CO$_2$ abundance on stellar carbon abundance. 
Including other commonly observed species such as the C$_2$ and CN radicals will be possible in future work, but we do not include them here as their parent species are not fully known \citep{Fray2005,Weiler2012}.

This model makes many approximations and assumptions. 
Furthermore, it only predicts a single ISO composition for each star, whereas in reality stars may produce ISOs with a range of compositions; this is due to both potential variation in the composition of a star's protoplanetary disk with position and time \citep[e.g.][]{Yunerman2026}, and differences in thermal processing \citep{Davidsson2021}.
Despite these simplifications, the \emph{trends} with parent star composition will be accurately captured: stars with a higher abundance of carbon will produce ISOs with more CO$_2$, and stars with a higher abundance of nitrogen will produce ISOs with more NH$_3$.

\section{Results}
\label{sec:results}

\begin{figure}
    \centering
    \includegraphics[width=\linewidth]{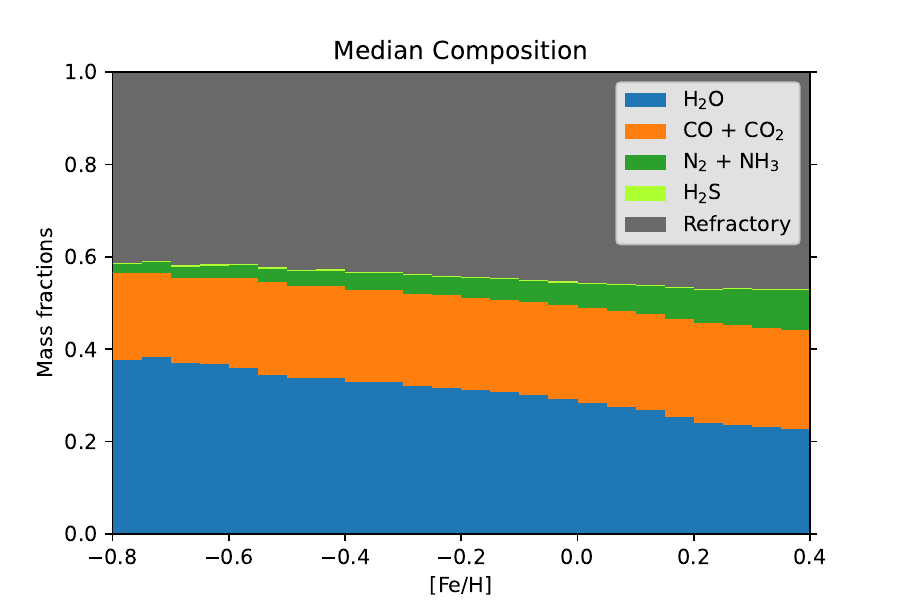}
    \includegraphics[width=\linewidth]{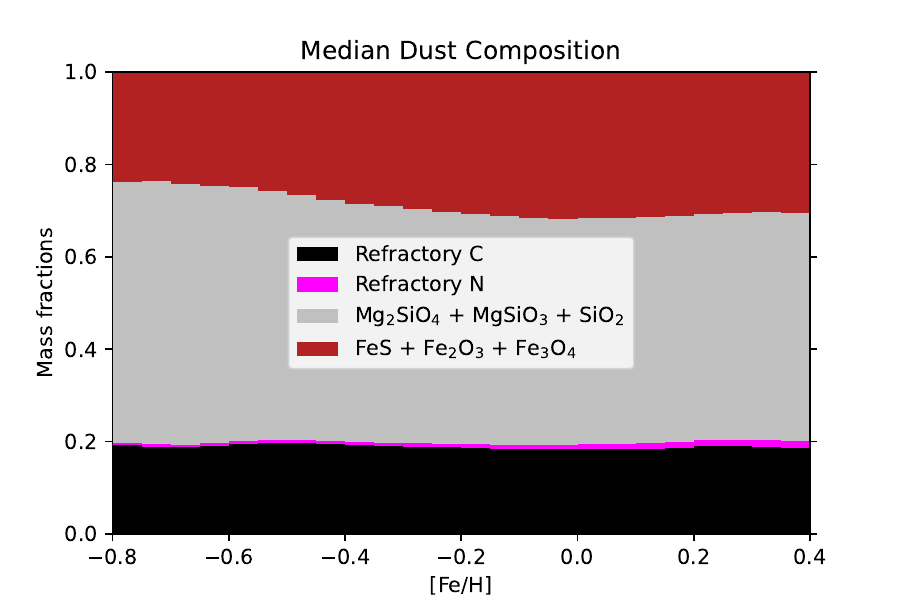}
    \caption{Median protoplanetary disk composition predicted by the chemical model at different metallicities [Fe/H]. \textbf{Top:} Median overall composition. \textbf{Bottom:} Median refractory composition.}
    \label{fig:massfractions}
\end{figure}
Applying our chemical model to our APOGEE stellar sample broadly predicts the compositions of ISOs in the Milky Way.
In Fig.~\ref{fig:massfractions} (top) we plot the median predicted protoplanetary disk composition at different metallicities. 
We have grouped species by element into carbon-based volatiles (CO and CO$_2$), nitrogen-based volatiles (N$_2$ and NH$_3$), hydrogen sulphide (H$_2$S),  water (H$_2$O), and refractory dust.

The most notable feature of the chemical model is that the mass fraction of H$_2$O in ISOs decreases with parent star [Fe/H].
This is expected from the chemistry in our model: H$_2$O can only form from oxygen left over after the formation of all other oxygen-containing molecules, and there is less left-over oxygen at higher metallicities.
This was the major trend noted in previous ISO composition work \citep[e.g.][]{Lintott2022,Hopkins2023,Hopkins2025b}; here we have extended our understanding to a fuller picture of the composition of ISOs.
Fig.~\ref{fig:massfractions} (bottom) shows how the median composition of the refractory material changes with parent star [Fe/H].

\begin{figure}
    \centering
    \includegraphics[width=\linewidth]{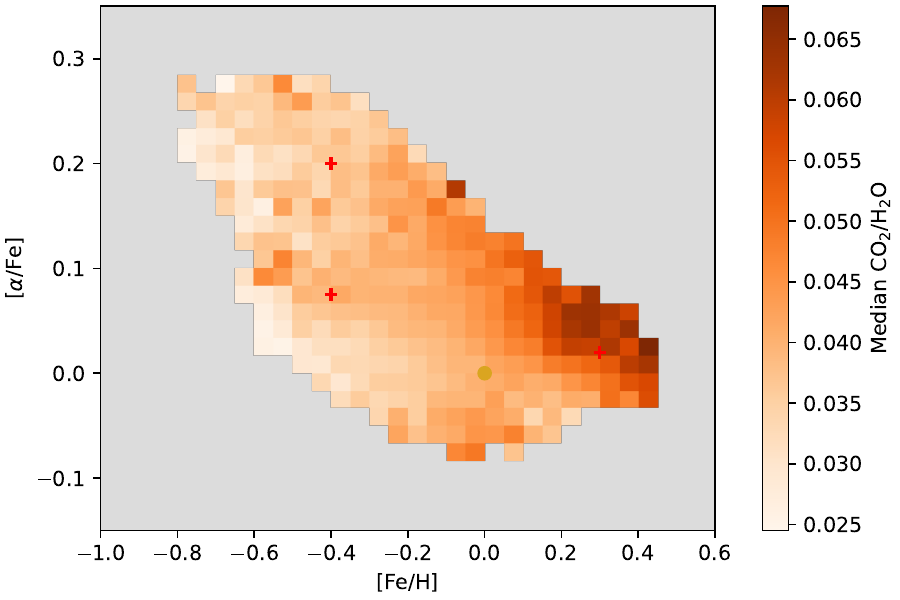}
    \includegraphics[width=\linewidth]{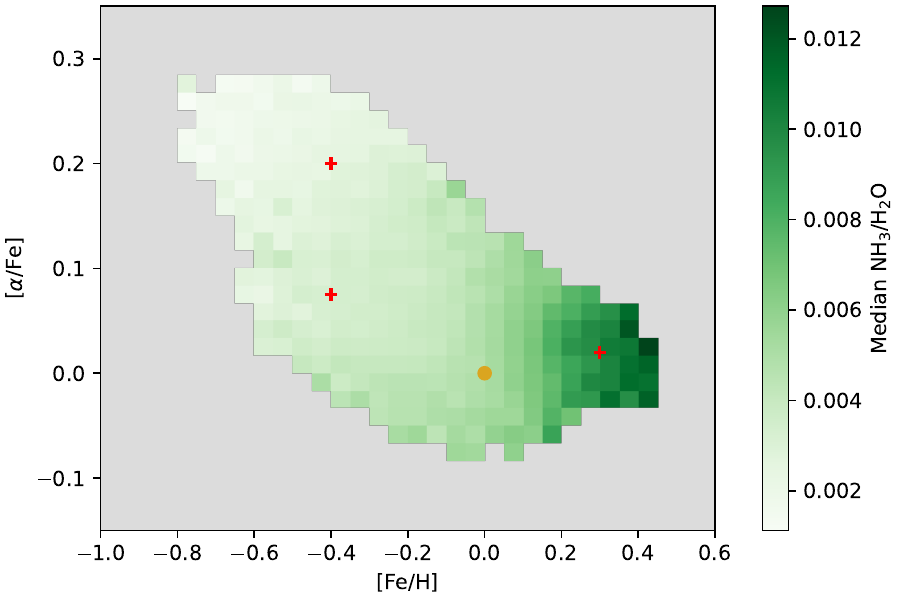}
    \includegraphics[width=\linewidth]{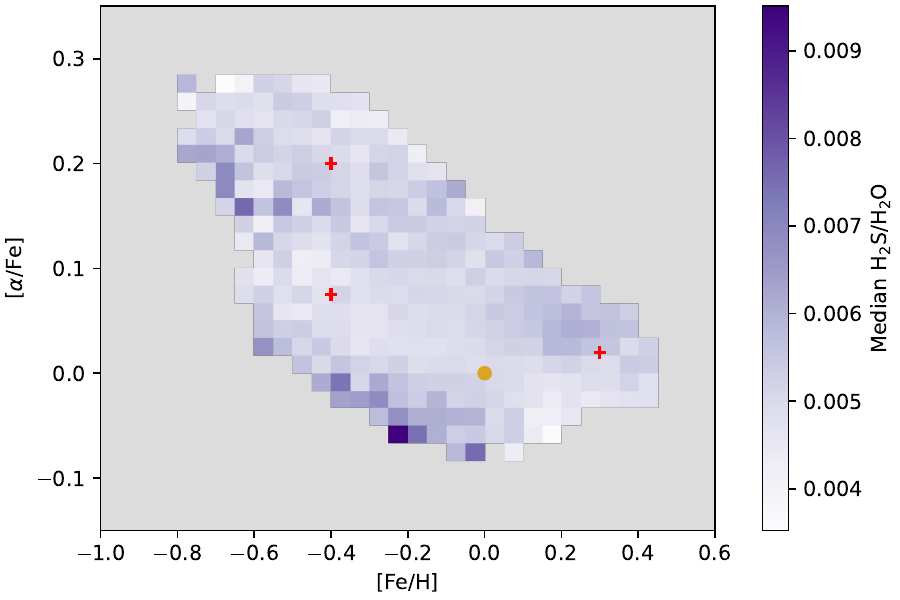}
    \caption{Median ISO abundances of CO$_2$, NH$_3$, and H$_2$S relative to H$_2$O across [$\alpha$/Fe]--[Fe/H] space. The gold circle marks the position of the Sun in this space and red crosses mark points at high [Fe/H], low [Fe/H], and high $\aFe$ at which the distribution of abundances is plotted in Fig~\ref{fig:aFeH_points}.}
    \label{fig:chem}
\end{figure}
As noted in \S\ref{sec:apogee}, the Milky Way's stellar populations are often displayed in [Fe/H]--[$\alpha$/Fe] because the elemental compositions, ages, and other properties of the Milky Way's stars show interesting correlations in this space.
We expect the Milky Way's ISOs to inherit similar correlations.
Fig \ref{fig:chem} thus shows how the median molecular abundances of CO$_2$, NH$_3$, and H$_2$S relative to H$_2$O varies between bins in [Fe/H]--[$\alpha$/Fe] space, including only bins with more than 20 stars.  
These three species have been chosen because they correlate with the stellar abundances of carbon, nitrogen and sulfur respectively, and are not hypervolatile. 
The median abundances ratios show strong trends with parent star chemical population, with higher-[Fe/H] stars producing more water-poor protoplanetary disks with higher ratios of CO$_2$/H$_2$O and NH$_3$/H$_2$O.
As expected from Fig.~\ref{fig:stars}, the relative abundance of NH$_3$ shows the strongest variation with stellar [Fe/H], meaning an ISO's ammonia abundance can be used to most easily infer the metallicity of its parent star.

\begin{figure}
    \centering
    \includegraphics[width=\linewidth]{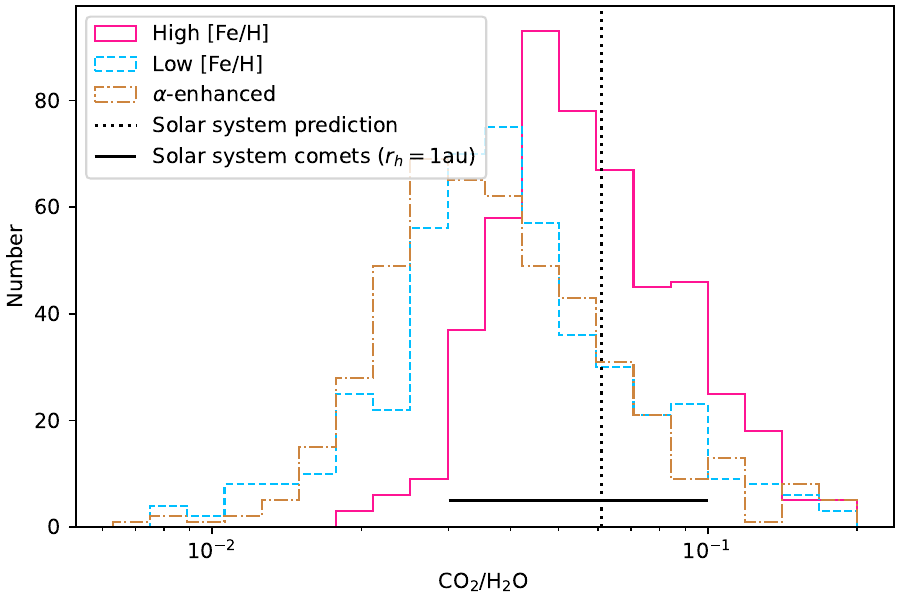}
    \includegraphics[width=\linewidth]{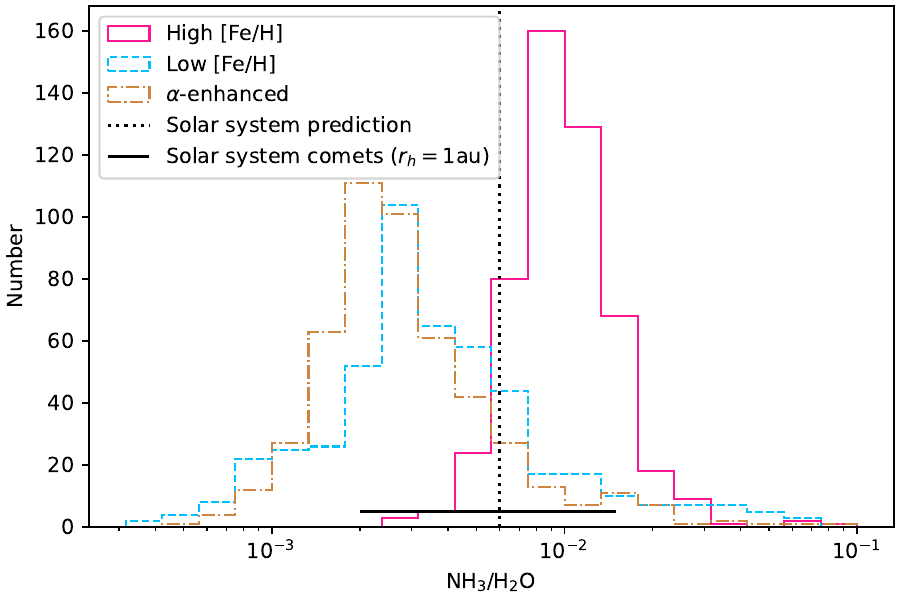}
    \includegraphics[width=\linewidth]{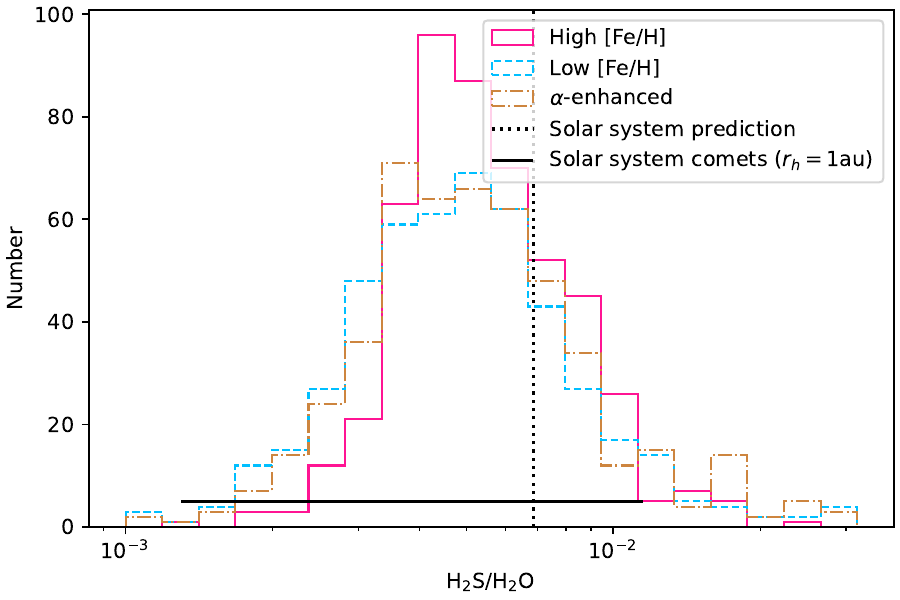}
    \caption{Plot of the distributions of ISO composition measures at each point in [$\alpha$/Fe]--[Fe/H] space marked on Figs.~\ref{fig:stars}\&\ref{fig:chem}, using a $k$NN method. A horizontal bar indicates the range of observed production rate ratios of Solar System comets at $r_h=1\,\mathrm{au}$, a proxy for their bulk compositions, and vertical lines mark the composition of Solar System comets predicted by our chemical model.}
    \label{fig:aFeH_points}
\end{figure}
We can then explore the spread of the predicted compositions of the ISOs, which is due to the large scatter in the elemental abundances of individual stars at each point in [Fe/H]--[$\alpha$/Fe] space.
Fig.~\ref{fig:aFeH_points} demonstrates this by plotting the distributions of CO$_2$/H$_2$O, NH$_3$/H$_2$O, and H$_2$S/H$_2$O at the three representative points in $\FeH$--$\aFe$ space marked with red crosses in Figures \ref{fig:stars} and \ref{fig:chem}:
\begin{itemize}
\item ``High [Fe/H]'' at $\FeH = 0.3,\,\,\,\aFe=0.02$
\item ``Low [Fe/H]'' at $\FeH = -0.4,\,\aFe=0.075$
\item ``$\alpha$-enhanced'' at $\FeH = -0.4,\,\aFe=0.2$
\end{itemize}
These distributions were calculated with a $k$-nearest neighbours method, taking the \(k=500\) nearest stars in [Fe/H]--[$\alpha$/Fe] space\footnote{Metric: \(d^2=(\Delta[\mathrm{Fe}/\mathrm{H}])^2+(\Delta[\alpha/\mathrm{Fe}])^2\)} in our APOGEE sample.

Fig.~\ref{fig:aFeH_points} shows that clear differences in composition should be expected between ISOs from stars in different parts of $\FeH$--$\aFe$ space, despite some overlap between the distributions.
In particular, the expected NH$_3$/H$_2$O ratios show little overlap between the high [Fe/H] and low [Fe/H] points, making the NH$_3$/H$_2$O ratio of an ISO a good tracer of its parent star metallicity.
Interestingly, our model predicts ISO H$_2$S/H$_2$O ratios to show little systematic variation with parent star [Fe/H] or [$\alpha$/Fe]; this may mean that complex sulfur chemistry sets the abundances of sulfur compounds observed in ISOs \citep[e.g.][]{Noonan2026} rather than the elemental sulfur abundances of their parent stars.

To aid comparison to Solar System comets, with vertical lines we have marked our model's prediction of the composition of Solar System comets using the Sun's photospheric elemental abundances \citep{Grevesse2007}.
We have also marked with horizontal bars the ranges of observed Solar System comet production rate ratios at heliocentric distance of $r_h=1\,\mathrm{au}$ (CO$_2$/H$_2$O, \citealt{HarringtonPinto2022}; NH$_3$/H$_2$O, \citealt{DelloRusso2016}; H$_2$S/H$_2$O, \citealt{Calmonte2016}).
This comparison is not exact, as the ratio of two species' production rates does not necessarily equal their ratio in the bulk composition of a comet's nucleus: comets are heterogeneous, and outgassing of different species occurs at different rates from different sublimation fronts and sources \citep{Sunshine2021,Davidsson2021}.
Additionally, a single comet's production rate ratios can vary strongly with heliocentric distance \citep{HarringtonPinto2022}, though production rate ratios at low heliocentric distance ($r_h\sim1\,\mathrm{au}$) are assumed to more closely match the bulk composition of a comet than production rate ratios measured further from the Sun.
It should be noted that 2I, 3I, and indeed most ISOs expected in upcoming discovery samples do not reach such low heliocentric distances \citep{Dorsey2025}, meaning their compositions must be inferred by comparing their measured production rates to Solar System comets observed at the same heliocentric distances (e.g. Appendix~\ref{sec:prodrates}).

Though our model only predicts one composition for each star in the APOGEE sample, the Solar System comet ranges plotted in Fig.~\ref{fig:aFeH_points} gives an indication of the range of compositions that a single star can produce.
Here, our predicted range exceeds the measured range, for CO$_2$/H$_2$O and NH$_3$/H$_2$O ratios.
This demonstrates that for an observed ISO, the dominant effect will be due to the elemental abundances of its parent star.
Finally, the broad agreement of our Solar predictions with the observed range of Solar System comet production rate ratios provides reassurance that our model is robust.

\section{Discussion}

\subsection{2I/Borisov and 3I/ATLAS}
\label{sec:2I3I}

The composition of an ISO can be inferred by comparing its production rate ratios to those of Solar System comets at the same heliocentric distance $r_h$.
We do this in Appendix~\ref{sec:prodrates}, finding that 2I/Borisov has a bulk NH$_3$/H$_2$O ratio similar to solar system comets, and 3I/ATLAS has a bulk NH$_3$/H$_2$O ratio approximately an order of magnitude lower than that of solar system comets.
Neither ISO has a well-constrained bulk CO$_2$/H$_2$O ratio: no CO$_2$ production rate was directly measured for 2I, and 3I's $Q(\mathrm{CO}_2)/Q(\mathrm{H}_2\mathrm{O})$ production rate ratio varied significantly, between a value similar to Solar System comets around its perihelion and a value significantly higher than Solar System comets at larger heliocentric distance, both before and after perihelion.
The trends predicted by our model mean the measured compositions of these ISOs can be used to make inferences about their other properties and the properties of their parent stars.

\begin{figure}
    \centering
    \includegraphics[width=\linewidth]{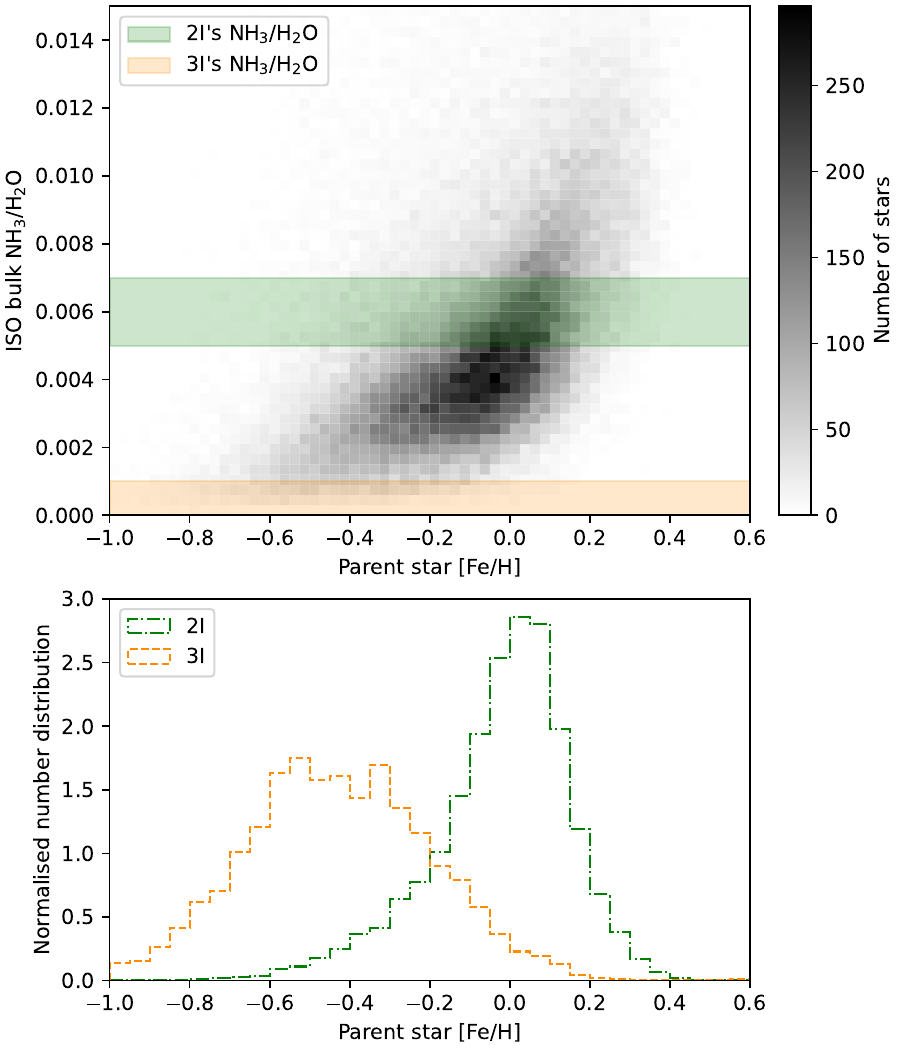}
    \caption{\textbf{Top:} Distribution of [Fe/H] and predicted ISO NH$_3$/H$_2$O for each star in  our stellar sample, with bands marking the approximate NH$_3$/H$_2$O of 2I and 3I. \textbf{Bottom:} Distribution of [Fe/H] of stars with predicted ISO NH$_3$/H$_2$O similar to 2I and 3I, indicating possible parent star metallicities for each.}
    \label{fig:nh3vsfeh}
\end{figure}
We can use the trend between a star's [Fe/H] and the NH$_3$/H$_2$O of its ISOs predicted by our chemical model to constrain the metallicity of the parent stars of 2I/Borisov and 3I/ATLAS.
Since ammonia production was not measured directly for either ISO, we take 2I's NH$_3$/H$_2$O as approximately the value predicted by our model for solar system comets, $\sim0.06$, and take 3I's NH$_3$/H$_2$O as $\lesssim0.001$.

As expected, Fig.~\ref{fig:nh3vsfeh} (top) shows predicted ISO NH$_3$/H$_2$O increases with increasing metallicity. 
Fig.~\ref{fig:nh3vsfeh} (bottom) shows the distribution of [Fe/H] for stars with predicted ISO NH$_3$/H$_2$O within 2I and 3I's approximate NH$_3$/H$_2$O ranges, providing estimates of the ranges of metallicities the parent stars of 2I and 3I may have had. 
Both are quite wide, indicating stars of a wide range of metallicities could have produced each ISO. 
2I's parent star likely falls within $-0.4\lesssim\FeH\lesssim0.3$, and 3I's parent star is likely lower metallicity than the Sun, falling within $-0.8\lesssim\FeH\lesssim0.0$.
These compositionally-inferred origins for 2I and 3I are compatible with the velocity-based inferences of \cite{Hopkins2025a} and \cite{Taylor2025}, and the inferences of \cite{Opitom2026} and \cite{Cordiner2026} using 3I's isotope ratios. 
We note that these results are not strongly dependant on the exact values chosen for the partition fractions of our chemical model; stars with higher [Fe/H] have higher elemental nitrogen abundance (Fig.~\ref{fig:stars}) and thus will produce ISOs of higher ammonia abundance.

\begin{figure}
    \centering
    \includegraphics[width=\linewidth]{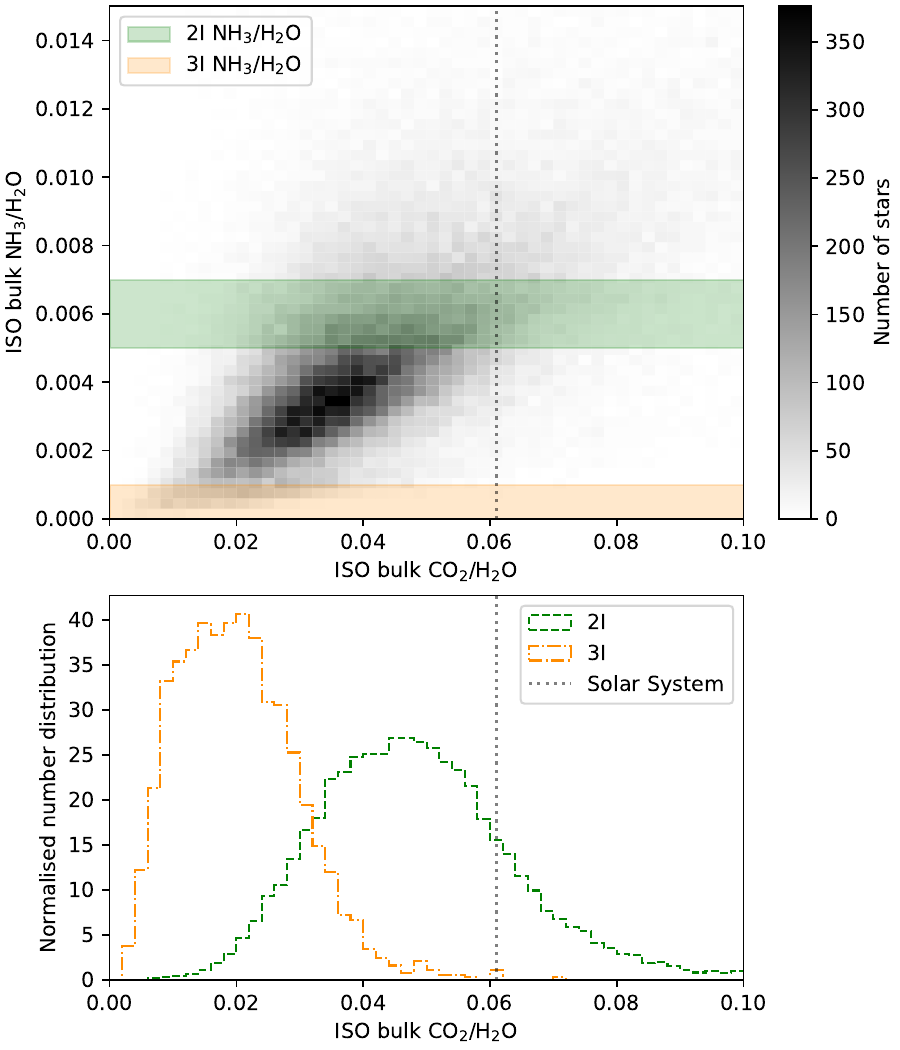}
    \caption{\textbf{Top:} Distribution of predicted ISO CO$_2$/H$_2$O with NH$_3$/H$_2$O for each star in our stellar sample, with bands marking the approximate NH$_3$/H$_2$O of 2I and 3I. \textbf{Bottom:} Distribution of predicted ISO NH$_3$/H$_2$O for stars with predicted ISO NH$_3$/H$_2$O similar to 2I and 3I, indicating possible bulk CO$_2$/H$_2$O ratio for each star.}
    \label{fig:nh3vsco2}
\end{figure}
Though neither 2I nor 3I's bulk CO$_2$/H$_2$O ratio is constrained, correlations between the molecular abundances predicted by our chemical model mean the bulk CO$_2$/H$_2$O ratios of 2I and 3I can be inferred from their bulk NH$_3$/H$_2$O ratios.
Fig.~\ref{fig:nh3vsco2} (top) shows the distribution of our stellar sample in the CO$_2$/H$_2$O and NH$_3$/H$_2$O of their ISOs as predicted by our chemical model; there is a clear positive correlation between the predicted bulk CO$_2$/H$_2$O and NH$_3$/H$_2$O ratios.

Fig.~\ref{fig:nh3vsco2} (bottom) shows the distribution of predicted bulk CO$_2$/H$_2$O ratios corresponding to NH$_3$/H$_2$O values within 2I and 3I's approximate NH$_3$/H$_2$O ranges. 
The distribution of possible CO$_2$/H$_2$O ratios for 2I is wide and extends either side of our predicted bulk composition for Solar System comets, so is not inconsistent with the indirect measurement of \cite{McKay2024} that 2I was CO$_2$ rich relative to Solar System comets.

The implied bulk CO$_2$/H$_2$O ratio for 3I/ATLAS is significantly lower than that of Solar System comets.
Interestingly, this prediction is inconsistent with observations, as 3I's measured $Q(\mathrm{CO}_2)/Q(\mathrm{H}_2\mathrm{O})$ production rate ratios are either similar to or higher than Solar System comets at the same heliocentric distances (Fig.~\ref{fig:carb-rh}, top). 
This discrepancy may be explained by Galactic cosmic ray processing conversion of CO to CO$_2$ \citep{Maggiolo2026}, but it is unclear if this mechanism could raise the CO$_2$ abundance of 3I by such a large factor while maintaining 3I's CO/H$_2$O abundance at levels similar to solar system comets and consistent with trapped hypervolatile CO.

\subsection{Hypervolatile-ice-rich ISOs}
\label{sec:hypervolatiles} 

Though the majority of volatile carbon and nitrogen in protoplanetary disks takes the form of CO and N$_2$, these hypervolatiles are only partially incorporated into typical comet nuclei as gases trapped in other ices due to their extremely low condensation temperatures. 
This is the case for both typical $\sim1\,\mathrm{km}$-diameter solar system comets \citep{Meech2009}, and appears to be the case for both 2I/Borisov and 3I/ATLAS too (Appendix \ref{sec:prodrates}). 
However, here we discuss the possible existence of hypervolatile-ice-rich ISOs.

ISOs with compositions including CO and N$_2$ ices are possible, and these hypervolatile-ice-rich ISOs may be common in the ISO population.
The majority of stars have lower masses and luminosities than the Sun \citep{Salpeter1955}.
M dwarfs with planetary systems can have luminosities as low as $5\times10^{-4}\,L_\odot$ \citep{Gillon2017}, and even-cooler brown dwarfs are observed with protoplanetary disks \citep{Damian2023}.
These protoplanetary disks could have midplane temperatures below the $20\,\mathrm{K}$ condensation temperature of CO and N$_2$ ice, and cometary bodies around these low-luminosity dwarfs would experience significantly less heating by insolation.
Some planetary systems will form with very little enrichment of the short-lived radionuclides thought to have heated forming comets in the Solar System \citep{Reiter2020,Forbes2021}, and ISO ejection by stellar flybys \citep{Pfalzner2021} will preferentially liberate ISOs on the cold edges of protoplanetary disks.
Once ejected, though stochastic heating episodes such as passing O-type stars and supernovae may heat some ISOs \citep{Stern2003}, C/2016 R2's retention of its hypervolatile ices through $4.5\,\mathrm{Gyr}$ in interstellar space show that avoiding these events is possible. 

Should they exist, we can use our model to predict the observable properties of hypervolatile-ice-rich ISOs.
Not limited by the amount of CO and N$_2$ that can be trapped in other ices, these ISOs would inherit the full composition of the protoplanetary disk, including the large abundances CO and N$_2$ have as the dominant forms of volatile carbon and nitrogen. 
This would give them significantly higher relative CO and N$_2$ production rates than Solar System comets, lying above the $Q(\mathrm{CO})/Q(\mathrm{H}_2\mathrm{O})$--$r_h$ trend of Fig.~\ref{fig:carb-rh} (bottom).
Their production rate ratios relative to water could be increased further by sublimation cooling by CO and N$_2$ ices \citep{Lisse2022}.
A hypervolatile-ice-rich ISO containing both CO and N$_2$ ice could also exhibit a higher coma N$^+_2$/CO$^+$ than even the N$_2$-rich Solar System comets of \cite{Anderson2023}, as it would not be limited by the poor trapping efficiency of N$_2$.
Though it is a \textit{bona fide} Solar System comet, C/2016 R2 may be a good analogue in its production rates for a hypervolatile-ice-rich ISO.

\subsection{Implications for Planetary Systems and the Galaxy}\label{sec:implications}

Though the sample of known cometary ISOs contains only two objects, the fact they originate around stars of very different metallicities has implications for planetary system formation.
Though there are strong correlations between the abundances of some types of exoplanets with host star metallicity \citep{Fischer2005,Buchhave2012}, it appears small planetesimals can both form and be ejected from planetary systems with a wide range of metallicities.

In particular, the low metallicity parent star of 3I/ATLAS has implications for the mechanism by which it was ejected.
The Solar System's giant planets ejected the bulk of the Sun's contributions to the ISO population \citep{Brasser2006} and ejection by giant planet is expected to be a common pathway for many ISOs \citep{Albrow2025}, but giant planets are rare around low-metallicity stars \citep{Johnson2010,Thorngren2016}.
This suggests 3I/ATLAS was ejected by another mechanism such as a stellar flyby \citep{Pfalzner2021}.
This explanation may be supported by 3I's isotope ratios \citep{Opitom2026,SalazarManzano2026} and coma N$^+_2$/CO$^+$ ratio \citep{Ferellec2026}, which imply 3I formed in the cold outer edges of its protoplanetary disk, where stellar flybys may preferentially liberate ISOs. 
Having just one example of an ISO ejected by a stellar flyby out of the three ISOs discovered so far suggests that ejection by stellar flyby is a common mechanism in the Galactic ISO population. 

As the number of known ISOs grows over the next decade, the metallicity of each parent star can be inferred from each ISO's ammonia abundance and the metallicity distribution of ISO parent stars can be measured and compared to the metallicity distribution of the solar neighbourhood. 
With this, the assumptions of \cite{Hopkins2023,Hopkins2025b} that higher metallicity stars will contribute more ISOs to the Galactic population can be tested.

These composition-based inferences about the properties of ISOs are highly complementary to velocity-based inferences such as those of \cite{Hopkins2025a} and \cite{Taylor2025}.
Whereas the composition-based method here can infer the metallicity of an ISO's parent star, a velocity-based method can provide an independent estimate of its age.
As the sample of known cometary ISOs grows, the combination of these two methods can be used to measure features of the Milky Way such as the age-metallicity relation \citep{Feltzing2001}.

\subsection{Future Observations}\label{sec:future}

The origins of 3I around an old, low-metallicity star were first inferred from its velocity \citep{Hopkins2025a,Taylor2025}, and then corroborated by measurements of its isotopic ratios \citep{Opitom2026,Cordiner2026}.
However, most ISOs discovered by LSST will be significantly less bright than 3I \citep{Dorsey2025} so similar isotopic measurements will be more difficult in ISO apparitions.
This means the method of inferring an ISO's origin from its chemical composition developed here may be the only method that can corroborate indirect velocity-based methods.
It is thus very important a full complement of production rates is measured, or at least constrained by upper limits, for 4I and beyond.

The ammonia abundance, measured with any of its photolytic daughter products $Q(\mathrm{NH}_x)$, is most important, as this allows the metallicity of the ISO's parent star to be inferred.
Next, $Q(\mathrm{CO})$ and $Q(\mathrm{N}_2)$ would allow the confirmation of a true hypervolatile-ice-rich ISO, if they exist.
$Q(\mathrm{CO}_2)$ is important as an abundant carbon-based volatile, not subject to the hypervolatility of CO, which we predict will correlate strongly with parent star carbon abundance, and may be sensitive to additional effects such as processing by Galactic cosmic rays \citep{Maggiolo2026}.
Finally, regular $Q(\mathrm{H}_2\mathrm{O})$ measurements through perihelion like \cite{Xing2020} and \cite{Combi2026} would allow other species to be compared on a common baseline.

\section{Conclusion}

The composition of an ISO we see passing through the inner Solar System reflects the conditions of its formation Gyr ago around a distant star, possibly very different from our own Sun.
Here we have developed a simple chemical model to predict the molecular compositions of cometary ISOs from the elemental abundances of their parent stars, and apply this to a stellar sample from the APOGEE survey to predict the distribution of ISO compositions in our Galaxy.
We find the ISO population of the Milky Way will show a wide range of compositions, wider than those of Solar System comets, with strong correlations between the molecular abundances of ISOs and the properties of their parent stars (Figs.~\ref{fig:chem}\&\ref{fig:aFeH_points}).
In particular, we predict that high metallicity stars will form ISOs rich in ammonia.

With these correlations we can make inferences about the origins of the two known cometary ISOs.
From their observed production rates it appears 2I/Borisov has a bulk NH$_3$/H$_2$O abundance similar to Solar System comets, whereas the NH$_3$/H$_2$O of 3I/ATLAS is approximately an order of magnitude lower. 
Using these estimates we can infer the range of possible parent star metallicities for each, finding 2I's parent star to lie within $-0.4\lesssim\FeH\lesssim0.3$, and 3I's parent star is likely lower metallicity than the Sun, falling within $-0.8\lesssim\FeH\lesssim0.0$ (Fig.~\ref{fig:nh3vsfeh}).

Using similar correlations between the predicted molecular ratios in ISOs, we show that 3I's low NH$_3$/H$_2$O abundance implies it should also have a CO$_2$/H$_2$O abundance lower than Solar System comets. 
This is inconsistent with its observed $Q(\mathrm{CO}_2)/Q(\mathrm{H}_2\mathrm{O})$ production rate ratios, which varied significantly but were never less than Solar System comets at the same heliocentric distance.
The excess CO$_2$ may be explained by an excess of Galactic cosmic ray processing during its time in interstellar space converting CO to CO$_2$ near its surface \citep{Maggiolo2026}, but is unclear if this is is consistent with 3I's near-Solar System $Q(\mathrm{CO})/Q(\mathrm{H}_2\mathrm{O})$ production rate ratios.

The comparative similarity of both 2I and 3I in their $Q(\mathrm{CO})/Q(\mathrm{H}_2\mathrm{O})$ ratios to Solar System comets at the same $r_h$ imply that, like Solar System comets, they both lack hypervolatile ices.
This is contrary to the interpretation of \cite{Bodewits2020} and \cite{Cordiner2020} who argue 2I did contain CO ice by equating the high $Q(\mathrm{CO})/Q(\mathrm{H}_2\mathrm{O})$ ratios they measure with the bulk CO/H$_2$O ratio in 2I's nucleus.
Despite the lack of hypervolatile ices in 2I/Borisov and 3I/ATLAS, we argue true hypervolatile-ice-rich ISOs are possible and may be common in the Galactic ISO population.
They would be distinguished from typical trapped-hypervolatiles-only comets by $Q(\mathrm{CO})/Q(\mathrm{H}_2\mathrm{O})$ production rate ratios significantly above the trend of Solar System comets with $r_h$ and an elevated coma N$_2^+$/CO$^+$ above the ratio expected from trapping, possibly resembling C/2016 R2 (PanSTARRS).

The three known ISOs are probably not representative of the whole Galaxy-spanning population, but the differences observed so far just between 1I/\okina Oumuamua, 2I/Borisov, and 3I/ATLAS hint at a highly-diverse population with origins in a wide range of planetary systems.
4I may be different yet again, but the model and predictions of this work provide a context in which it can be placed, shedding light on its distant origins.  
With observations of common molecules we can constrain the origins of even faint ISOs, an exciting prospect for future surveys.

\begin{acknowledgments}
We thank
Sven Buder and Ben Lowe for advice about the GALAH survey, and
Cyrielle Opitom, Lea Ferellec, Matthew Belyakov, Dennis Bodewits, Sarah Anderson, Joe Masiero and Colin Snodgrass for interesting and informative discussions at ACM2026.
MJH appreciates support from the Elaine P. Snowden Fellowship.
M.T.B. appreciates support by the Rutherford Discovery Fellowships from New Zealand Government funding, administered by the Royal Society Te Ap\={a}rangi.
CJL thanks Breakthrough Listen for support.

\end{acknowledgments}

\appendix\restartappendixnumbering

\section{Production Rate Ratios}\label{sec:prodrates}
Here we summarise measured production rate ratios for 2I and 3I, and compare them to observations of solar system comets.
Though the production rate ratio of two species does not necessarily equal their ratio in the bulk composition of a comet's nucleus, especially at larger heliocentric distances, they should be correlated.
Thus an ISO's composition can be inferred by comparing its measured production rate ratios to those of solar system comets at the same heliocentric distance. 

\begin{figure*}
    \centering
    \includegraphics[width=\linewidth]{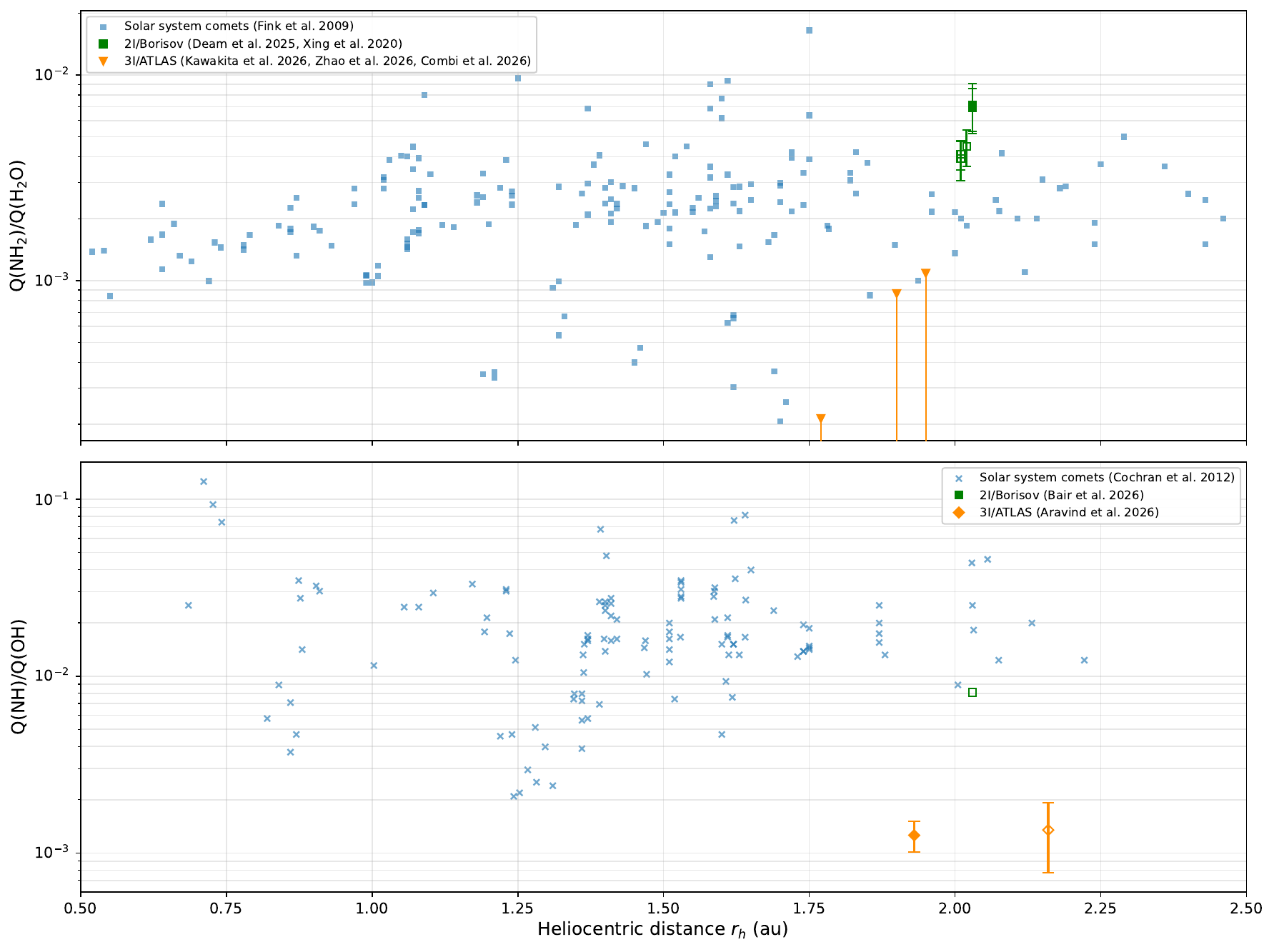}
    \caption{Production rate ratios $Q(\mathrm{NH}_2)/Q(\mathrm{H}_2\mathrm{O})$ (top) and $Q(\mathrm{NH})/Q(\mathrm{OH})$ (bottom) of 2I/Borisov and 3I/ATLAS compared to Solar System comets. 
    The markers for 2I and 3I are unfilled for pre-perihelion observations and filled post-perihelion. 
    The low ammonia content of 3I indicates formation around a low-metallicity star.}
    \label{fig:ammo-rh}
\end{figure*}
In this work we have shown an ISO's bulk NH$_3$/H$_2$O composition correlates strongly with the metallicity of its parent star.
Though ammonia can be detected in a comet's coma directly with high-resolution near-infrared spectroscopy \citep[e.g.][]{Lippi2021}, no direct NH$_3$ production rates have been reported for 2I or 3I.
Instead, in Fig.~\ref{fig:ammo-rh} we plot as proxies $Q(\mathrm{NH}_2)/Q(\mathrm{H}_2\mathrm{O})$ (top) and $Q(\mathrm{NH})/Q(\mathrm{OH})$ (bottom) production rate ratios for 2I/Borisov, 3I/ATLAS, and Solar System comets \citep{Fink2009,Cochran2012}.
We use OH production as a proxy for water production with NH because the two are simultaneously detectable in the same near-UV range, but use H$_2$O to compare NH$_2$ production rates as a common point of comparison. 

No simultaneous measurements of $Q(\mathrm{NH}_2)/Q(\mathrm{H}_2\mathrm{O})$ exist for 2I and 3I, so we calculate this from near-simultaneous measurements from different instruments. 
For 2I we divided the absolute $Q(\mathrm{NH}_2)$ values by \cite{Deam2025} by the absolute values of $Q(\mathrm{H}_2\mathrm{O})$ by \cite{Xing2020} where the measurements of the two species were taken within three days of each other. 
For 3I, we multiplied the relative upper limits on the $Q(\mathrm{NH}_2)/Q(\mathrm{C}_2)$ production rate ratios by \cite{Kawakita2026} by the absolute $Q(\mathrm{C}_2)$ production rates by \cite{Zhao2026} and divided by the absolute $Q(\mathrm{H}_2\mathrm{O})$ by \cite{Combi2026} where the measurements were taken within three days of each other.
Simultaneous measurements of $Q(\mathrm{NH})/Q(\mathrm{OH})$ are reported for both 2I/Borisov \citep{Bair2026} and 3I/ATLAS \citep{Aravind2026}.

2I/Borisov has a $Q(\mathrm{NH}_2)/Q(\mathrm{H}_2\mathrm{O})$ ratio at the high end of the range of Solar System comets at $r_h=2\,\mathrm{au}$, and a $Q(\mathrm{NH})/Q(\mathrm{OH})$ at the lower end of the Solar System range.
We interpret this as an approximately Solar System ammonia abundance, thus implying 2I's parent star likely had a near-solar metallicity.
Conversely, the $Q(\mathrm{NH}_2)/Q(\mathrm{H}_2\mathrm{O})$ upper limits and $Q(\mathrm{NH})/Q(\mathrm{OH})$ measurements on 3I/ATLAS show it to be highly depleted in ammonia by approximately an order of magnitude, implying it to have originated around a low-metallicity star.

\begin{figure*}
    \centering
    \includegraphics[width=\linewidth]{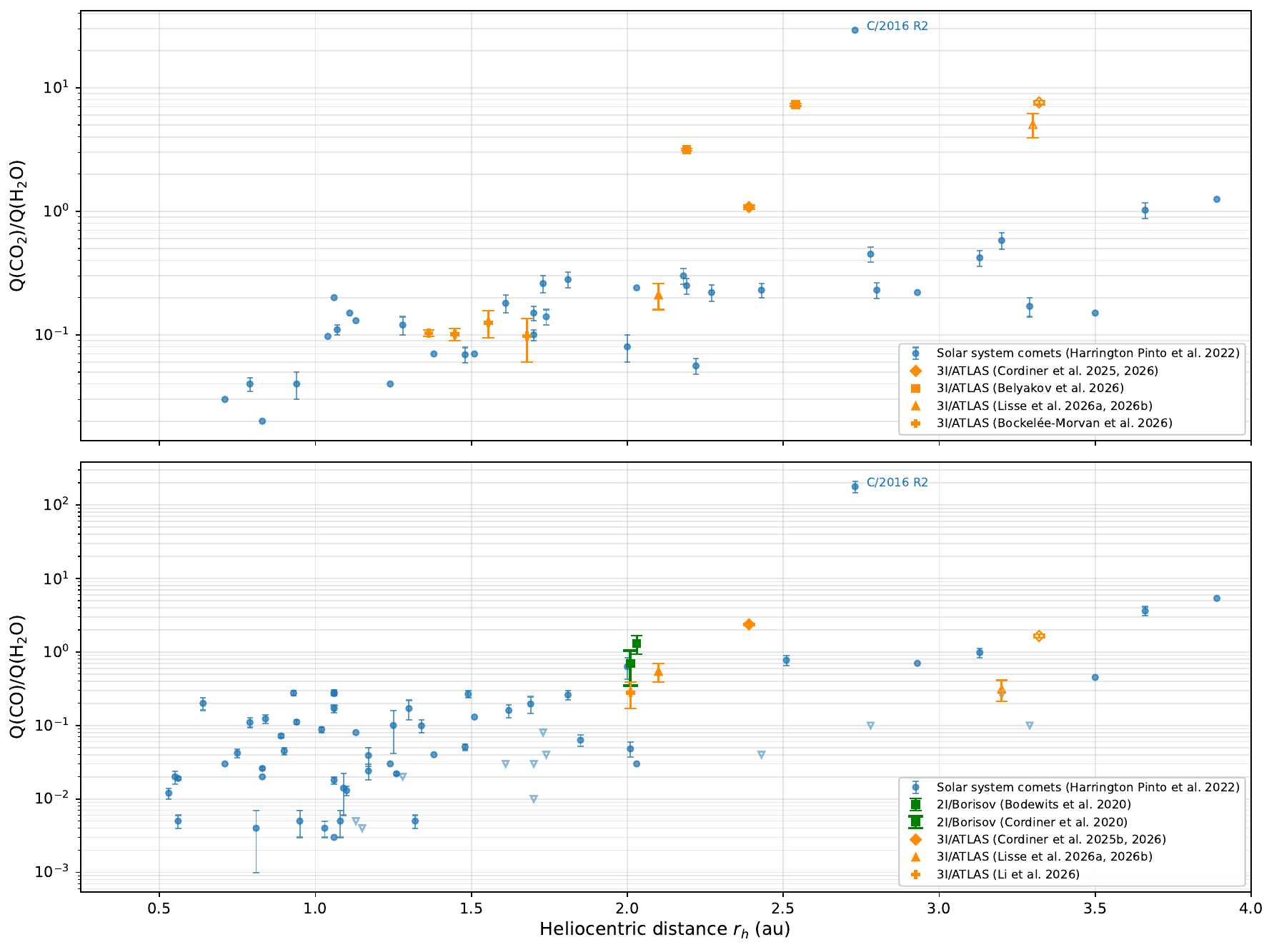}
    \caption{Production rate ratios $Q(\mathrm{CO}_2)/Q(\mathrm{H}_2\mathrm{O})$ (top) and $Q(\mathrm{CO})/Q(\mathrm{H}_2\mathrm{O})$ (bottom) of 2I/Borisov and 3I/ATLAS compared to Solar System comets. The markers for 2I and 3I are unfilled pre-perihelion and filled post-perihelion.  3I's $Q(\mathrm{CO}_2)/Q(\mathrm{H}_2\mathrm{O})$ lies above the trend line of Solar System comets at high $r_h$ but falls amid the scatter around perihelion, complicating its interpretation. Both ISOs show a $Q(\mathrm{CO}_2)/Q(\mathrm{H}_2\mathrm{O})$ near the trend of Solar System comets, implying both also lack hypervolatile CO ice.}
    \label{fig:carb-rh}
\end{figure*}
Next we consider the relative production rates of carbon volatiles. 
In Fig.~\ref{fig:carb-rh} we plot the $Q(\mathrm{CO}_2)/Q(\mathrm{H}_2\mathrm{O})$ (top) and $Q(\mathrm{CO})/Q(\mathrm{H}_2\mathrm{O})$ (bottom) production rate ratios for 2I/Borisov, 3I/ATLAS, and Solar System comets \citep{HarringtonPinto2022}.
Solar System comets show a clear trend, with both $Q(\mathrm{CO}_2)/Q(\mathrm{H}_2\mathrm{O})$ and $Q(\mathrm{CO}_2)/Q(\mathrm{H}_2\mathrm{O})$ increasing steadily with heliocentric distance.
The ratios plotted for 2I and 3I are all reported directly by \cite{Cordiner2020,Bodewits2020,Cordiner2025,Cordiner2026,Lisse2026a,Lisse2026b,Belyakov2026} and \cite{Bockelee-Morvan2026}.

Due to a lack of operational space-based infrared facilities during 2I's apparition, no CO$_2$ production rate was directly measurable, though \cite{McKay2024} inferred its $Q(\mathrm{CO}_2)/Q(\mathrm{H}_2\mathrm{O})$ ratio was high relative to Solar System comets from its [OI] emission.
3I's measured $Q(\mathrm{CO}_2)/Q(\mathrm{H}_2\mathrm{O})$ ratios (Fig.~\ref{fig:carb-rh}, top) clearly lie above the trend of the Solar System comets for $r_h>2\,\mathrm{au}$ both before and after perihelion, however observations closer to its perihelion ($q=1.36\,\mathrm{au}$) have significantly lower $Q(\mathrm{CO}_2)/Q(\mathrm{H}_2\mathrm{O})$, within the Solar System comet scatter.
This makes it difficult to constrain 3I's bulk CO$_2$/H$_2$O, although it appears to be either similar to or higher than Solar System comet compositions, but not lower. 

Finally we consider the carbon monoxide production rate ratios (Fig.~\ref{fig:carb-rh}, bottom). 
The $Q(\mathrm{CO})/Q(\mathrm{H}_2\mathrm{O})$ production rate ratios for both 2I/Borisov and 3I/ATLAS, unlike the other production rate ratios, lie within or just outside the scatter of solar system comets.
In fact, the single outlier in $Q(\mathrm{CO})/Q(\mathrm{H}_2\mathrm{O})$ is C/2016 R2 (PanSTARRS), a \textit{bona fide} Solar System comet. 
This comparative similarity of 2I and 3I to typical Solar System comets is due to the nature of CO as a hypervolatile:
Due to its incredibly low condensation temperature ($\sim20\,\mathrm{K}$) CO is only present in typical, $\sim1\,\mathrm{km}$-diameter Solar System comets as a volatile trapped in other ices \citep{Meech2009}, not as pure CO ice.
In this case, the relative abundance of CO is limited by the trapping efficiency of CO and does not equal the abundance of CO in the protoplanetary disk. 
Consequently, the comparative similarity of 2I/Borisov and 3I/ATLAS in $Q(\mathrm{CO})/Q(\mathrm{H}_2\mathrm{O})$ to typical Solar System comets despite origins in different protoplanetary disks implies that they also lack hypervolatile ices.
For 2I/Borisov this is contrary to \cite{Bodewits2020} and \cite{Cordiner2020} who argue 2I did contain CO ice by equating the high $Q(\mathrm{CO})/Q(\mathrm{H}_2\mathrm{O})$ ratios they measure with the bulk CO/H$_2$O ratio in 2I's nucleus.
For 3I/ATLAS, a composition containing only trapped hypervolatiles is corroborated by \cite{Ferellec2026}, who find 3I to have a similar coma N$_2^+$/CO$^+$ ratio to Solar System comets and therefore consistent with trapped hypervolatiles.
It thus appears that the two cometary ISOs discovered so far contain only trapped hypervolatiles, like typical Solar System comets.

\bibliography{export-bibtex}
\bibliographystyle{aasjournal}

\end{document}